\documentclass[aps, prl, superscriptaddress, notitlepage, reprint]{revtex4-1}
\usepackage[english]{babel}
\usepackage{amsmath,amsthm}
\usepackage{amsfonts}
\usepackage[pdfborder={0 0 0}, colorlinks=true, urlcolor=blue, linkcolor=blue, citecolor=blue]{hyperref}
\usepackage{color}
\usepackage{graphicx}
\usepackage{float}
\usepackage{subfigure}
\usepackage{lipsum}

\theoremstyle{definition}

\theoremstyle{remark}

\begin{document}

\title{Enhanced entanglement and asymmetric EPR steering between magnons}
\author{Sha-Sha Zheng}
\affiliation{State Key Laboratory for Mesoscopic Physics and Frontiers Science Center for Nano-Optoelectronics, School of Physics $\&$ Collaborative Innovation Center of Quantum Matter, Peking University, Beijing 100871, China}
\affiliation{Beijing Academy of Quantum Information Sciences, Haidian District, Beijing 100193, China}
\affiliation{Collaborative Innovation Center of Extreme Optics, Shanxi University, Shanxi 030006, China}
\author{Feng-Xiao Sun}
\affiliation{State Key Laboratory for Mesoscopic Physics and Frontiers Science Center for Nano-Optoelectronics, School of Physics $\&$ Collaborative Innovation Center of Quantum Matter, Peking University, Beijing 100871, China}
\affiliation{Beijing Academy of Quantum Information Sciences, Haidian District, Beijing 100193, China}
\affiliation{Collaborative Innovation Center of Extreme Optics, Shanxi University, Shanxi 030006, China}
\author{Huai-Yang Yuan}
\affiliation{Institute for Theoretical Physics, Utrecht University, 3584CC Utrecht, The Netherlands}
\author{Zbigniew Ficek}
\affiliation{Quantum Optics and Engineering Division, Institute of Physics, University of Zielona G\'{o}ra, Szafrana 4a, Zielona G\'{o}ra 65-516, Poland}
\author{Qi-Huang Gong}
\affiliation{State Key Laboratory for Mesoscopic Physics and Frontiers Science Center for Nano-Optoelectronics, School of Physics $\&$ Collaborative Innovation Center of Quantum Matter, Peking University, Beijing 100871, China}
\affiliation{Beijing Academy of Quantum Information Sciences, Haidian District, Beijing 100193, China}
\affiliation{Collaborative Innovation Center of Extreme Optics, Shanxi University, Shanxi 030006, China}
\author{Qiong-Yi He}
\email[Electronic address: ]{qiongyihe@pku.edu.cn}
\affiliation{State Key Laboratory for Mesoscopic Physics and Frontiers Science Center for Nano-Optoelectronics, School of Physics $\&$ Collaborative Innovation Center of Quantum Matter, Peking University, Beijing 100871, China}
\affiliation{Beijing Academy of Quantum Information Sciences, Haidian District, Beijing 100193, China}
\affiliation{Collaborative Innovation Center of Extreme Optics, Shanxi University, Shanxi 030006, China}

\begin{abstract}

The generation and manipulation of strong entanglement and Einstein-Podolsky-Rosen (EPR) steering in macroscopic systems are outstanding challenges in modern physics. Especially, the observation of asymmetric EPR steering is important for both its fundamental role in interpreting the nature of quantum mechanics and its application as resource for the tasks where the levels of trust at different parties are highly asymmetric. Here, we study the entanglement and EPR steering between two macroscopic magnons in a hybrid ferrimagnet-light system. In the absence of light, the two types of magnons on the two sublattices can be entangled, but no quantum steering occurs when they are damped with the same rates. In the presence of the cavity field, the entanglement can be significantly enhanced, and strong two-way asymmetric quantum steering appears between two magnons with equal dispassion. This is very different from the conventional protocols to produce asymmetric steering by imposing additional unbalanced losses or noises on the two parties at the cost of reducing steerability. The essential physics is well understood by the unbalanced population of acoustic and optical magnons under the cooling effect of cavity photons. Our finding may provide a novel platform to manipulate the quantum steering and the detection of bi-party steering provides a knob to probe the magnetic damping on each sublattice of a magnet. \\


\textbf{PACS number(s): 03.67.-a, 03.65.Ud, 03.67.Bg}\\

\end{abstract}
\maketitle

\section{1 Introduction}

Hybridizing two or more quantum systems, aiming at combining complementary functionalities of distinct systems to obtain multitasking capabilities, is very necessary and crucial for implementing quantum information processing~\cite{Lachance2019}. Recently, significant attentions have been paid to hybrid quantum systems involving magnons, which characterize a collective motion of a large number of spins in a macroscopic magnetically ordered materials. It has been both theoretically and experimentally confirmed that the magnons can be coherently coupled with various physical systems, including superconducting qubits~\cite{Tabuchi2015}, phonons~\cite{ZhangXufeng2016}, microwave photons~\cite{Soykal2010,Hubel2013,ZhangXufeng2014,Tabuchi2014}, and optical photons~\cite{Haigh2016,Osada2018}. In addition, the magnons possess the distinguished advantage of the long lifetime, high spin density, and easy tunability, providing a promising experimental platform to achieve strong light-matter interaction. Particularly, the coupling between magnons with the microwave cavity photons can easily reach strong-coupling regime~\cite{Hubel2013,ZhangXufeng2014,Tabuchi2014} even ultrastrong-coupling regime~\cite{Goryachev2014,Bourhill2016,Kostylev2016}, leading to cavity-magnon polaritons. The photon-magnon polariton provides great opportunities to study many novel phenomena, such as level attraction~\cite{Harder2018,YuWeiChao2019}, exceptional surface~\cite{ZhangXufeng2019ES}, bistability~\cite{WangYipu2018}, nonreciprocity~\cite{WangYipu2019}, magnon blockade~\cite{LiuZengXing2019}, etc. Furthermore, hybrid quantum systems based on magnons may pave the way for achieving microwave-to-optical quantum transducers~\cite{Hisatomi2016}, memories~\cite{ZhangXufeng2015}, magnon-based data processing and computing circuits~\cite{Chumak2015}, and ultra-sensitive detection~\cite{Lachance2019} in the area of quantum information science and engineering.

Quantum entanglement, the key resource for quantum information, has been extensively studied from the fundamental perspective and the applications ranging from quantum cryptography, quantum computation to quantum metrology~\cite{Horodecki2009}. Considerable attention has been devoted to quantum steering~\cite{Reid2009,SteeringReview2020}, a strict subset of quantum entanglement. The concept of quantum steering was originally introduced by Schr\"{o}dinger~\cite{Schrodinger1935} in response to the famous Einstein-Podolsky-Rosen (EPR) paradox~\cite{Einstein1935,Reid1989}, and it describes the ability of one system to remotely adjust (steer) the state of the other entangled system by local measurements. The presence of EPR steering enables verification of shared entanglement even when one party's measurements are untrusted~\cite{Wiseman2007}. This provides extra security to various information protocols, such as one-sided device-independent quantum cryptography and related protocols~\cite{Walk2016,Armstrong2015,He2015,Li2015}. In addition, the research about the quantum correlations between two massive objects is extremely significant from the fundamental perspective in terms of quantum-to-classical crossover~\cite{Zurek2003review}, macroscopic quantum effects~\cite{Florian2018}, wave-function collapse theory~\cite{Angelo2013}. Owing to the large size of the magnon system, hybrid magnonical systems have emerged as a promising platform to study macroscopic quantum phenomenon. Recently, several schemes have been proposed to generate different types of quantum correlations between magnons with photons or phonons~\cite{Yuan2020PRL,LiJie2018,ZhangZhedong2019,Yuan2020PRB}. However, whether asymmetric EPR steering can be achieved in such hybrid systems is not known yet.

In this paper, we investigate the quantum entanglement and EPR steering of two macroscopic magnons in a two-sublattice ferrimagnetic system coupled to a microwave cavity. The magnons are directly coupled to each other due to the exchange interaction between spins on the two sublattices. We show that magnons can be entangled due to the exchange coupling, while the magnon modes in ferrimagnet can be stronger entangled than in the antiferromagnet in the presence of cavity field. Moreover, strong asymmetric two-way steering between the magnons is generated even if the modes are damped with the same rates. This is in contrast to the case when the cavity is absent, in which one-way steering may occur only when the magnons decay with different rates. Moreover, a stronger steerability can be produced in antiferromagnet than in ferrimagnet. These results are shown to be sensitive to the cooling of the magnon modes by the cavity mode that the cavity plays a role of a cold bath which cools the magnons toward their vacuum state. We find that the cooling effect disappears when one of the magnon modes and the cavity mode are mutually perfectly coherent. As a result, a dip occurs in the enhanced entanglement at which the degrees of entanglement and steering are equal to those obtained in the absence of the cavity. We explain this effect as resulting from the perfect destructive interference between two channels the magnon mode and the photon mode are coupled. 

\section{2 Model and method}\label{Method}
\begin{figure}
\includegraphics[width=0.7\columnwidth]{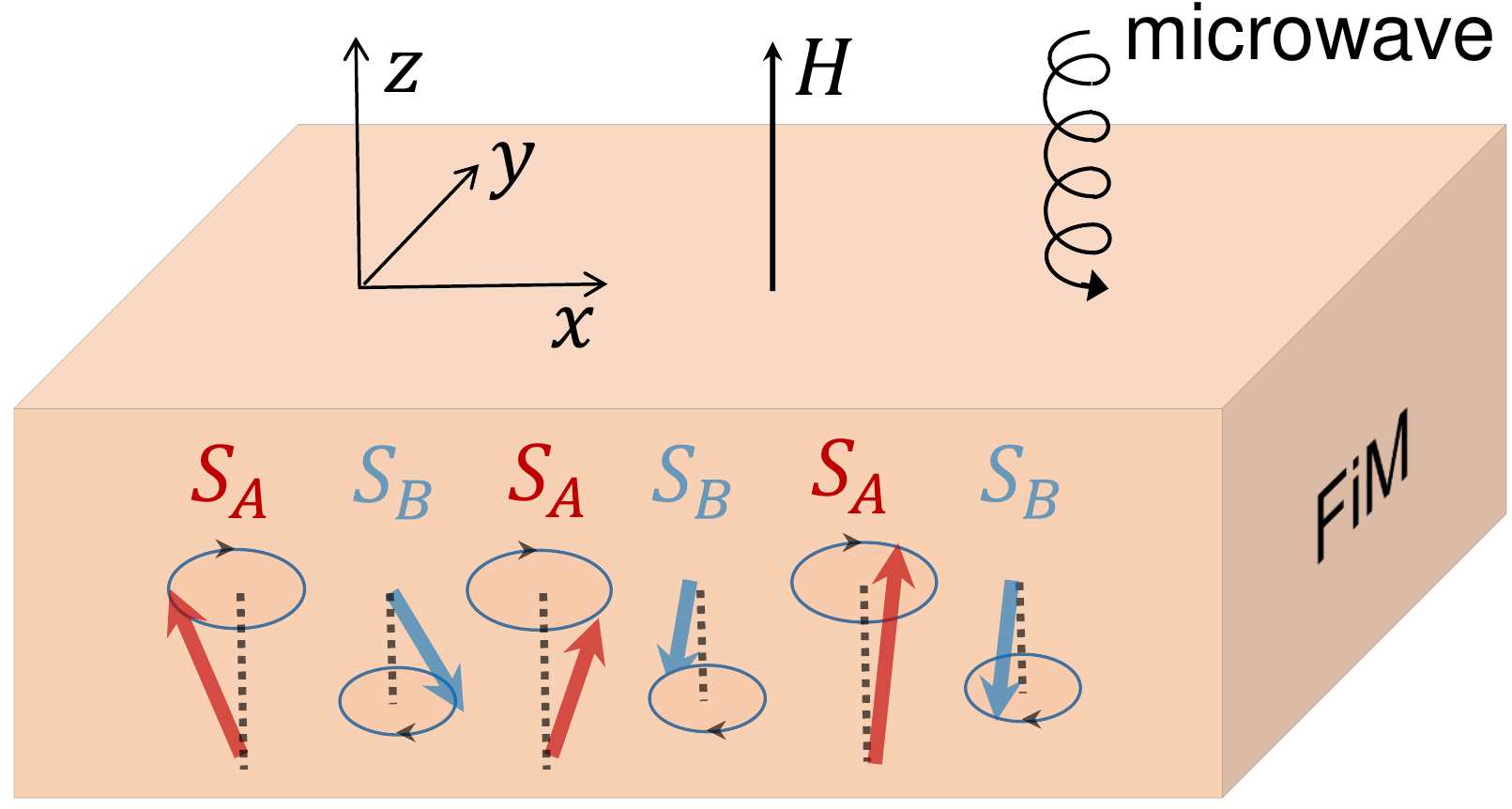}
\caption{Schematic illustration of a two-sublattice ferrimagnet coupled with a circularly polarized electromagnetic wave. The red and blue arrows represent the spins on the sublattice $A$ and $B$, respectively. $H$ pointing along the $z$ direction is the external static field.}
\label{fig1}
\end{figure}

We consider a general model composed of a circularly polarized microwave cavity and a ferrimagnet consisting of two sublattices $A$ and $B$, whose spins in each sublattice are oriented in opposite directions along the $z$ axis, as illustrated in Fig.~\ref{fig1}. The Hamiltonian of this system is described by~\cite{YuanApl}
\begin{equation}\label{H}
\mathcal{H}=\mathcal{H}_{\mathrm{FiM}}+\mathcal{H}_c+\mathcal{H}_{\mathrm{int}},
\end{equation}
where $\mathcal{H}_{\mathrm{FiM}}$ represents the Hamiltonian for ferrimagnets, $\mathcal{H}_c$ denotes the cavity mode, and $\mathcal{H}_{\mathrm{int}}$ is the term describing the magnet-light interaction. These three terms take the following detailed form
\begin{eqnarray}
&&\mathcal{H}_{\mathrm{FiM}}=J \sum_{l,\delta}  (\mathbf{S}_{A}^l \cdot \mathbf{S}_{B}^{l+\delta}+ \mathbf{S}_{B}^l \cdot \mathbf{S}_{A}^{l+\delta})+\mathcal{H}_A+ \mathcal{H}_B,\nonumber\\
&&\mathcal{H}_c=\frac{1}{2}\int \mathrm{d}v \left ( \epsilon_0 \mathbf{E}_c^2
+ \mu_0\mathbf{H}_c^2 \right ),\nonumber\\
&&\mathcal{H}_{\mathrm{int}}=-\sum_{l} (g_A\mu_B\mathbf{S}_{A}^l+g_B\mu_B\mathbf{S}_{B}^l)\cdot \mathbf{H}_c,
\end{eqnarray}
where $\mathbf{S}_{A}^l$ and $\mathbf{S}_{B}^l$ describe the spin on site $l$ of two sublattices $A$ and $B$, respectively, $\delta$ represents the displacement of two nearest spins, and $J>0$ is the nearest-neighbor exchange integral. The last two terms of $\mathcal{H}_{\mathrm{FiM}}$ come from the contributions of the external static field $\mathbf{H}$ and anisotropy field $\mathbf{H}_{\mathrm{an,}i}~(i=A,B)$, respectively, which read $\mathcal{H}_A=-\sum_l g_A\mu_B(\mathbf{H}+\mathbf{H}_{\mathrm{an},A}) \cdot \mathbf{S}_{A}^l$ and $\mathcal{H}_B=-\sum_l g_B\mu_B(\mathbf{H}+\mathbf{H}_{\mathrm{an},B}) \cdot \mathbf{S}_{B}^l$. The coefficients $g_i~(i=A, B)$ and $\mu_B$ are effective g-factor for sublattice $i$ and the Bohr magneton. $\mathbf{E}_c$ and $\mathbf{H}_c$ are respectively the electric and magnetic field vectors of the electromagnetic (EM) wave inside the cavity, and $\epsilon_0$, $\mu_0$ are the permittivity and susceptibility in free space, respectively.

To investigate the theory with the Hamiltonian~(\ref{H}), it is convenient to introduce bosonic photon operators for the EM wave inside the cavity and magnon operators for spin waves using Holstein-Primakoff transformation~\cite{Holstein1940}. Then the Hamiltonian is expressed as
\begin{align}
\mathcal{H}&=  \omega_a a^\dagger a + \omega_b b^\dagger b +g_{ab}(a^\dagger
b^\dagger + a b) \nonumber\\
&\ \ \ \  + \omega_c c^\dagger c + g_{ac} \left ( a^ \dagger c^\dagger + a c \right)
+ g_{bc} \left (  b^\dagger c + b c^\dagger \right) ,
\label{H_bosonic}
\end{align}
where $a, a^\dagger$ ($b, b^\dagger$) represent the annihilation, creation operators for the sublattice $A$ ($B$), respectively, and $c, c^\dagger$ correspond to photons inside the cavity. The photon frequency is $\omega_c$, and the frequency of two magnons is $\omega_i=\omega_{i,\mathrm{ex}}+\omega_{i,\mathrm{an}}+\omega_{i,0}$ with coefficients $\omega_{a,\mathrm{ex}}=2ZS_BJ,~\omega_{b,\mathrm{ex}}=2ZS_AJ,~\omega_{a,\mathrm{an}}=g_A\mu_BH_{\mathrm{an},A},~\omega_{b,\mathrm{an}}=g_B\mu_BH_{\mathrm{an},B},~\omega_{a,0}=g_A\mu_BH,~\omega_{b,0}=-g_B\mu_BH$, in which $Z$ denotes the coordination number. $g_{ab}=2\gamma_kZ\sqrt{S_A S_B}J$ indicates the coupling strength between two magnon modes, where $\gamma_k=Z^{-1}\sum_{\delta}e^{ik\delta}$ describes the structure factor. The coupling strengths between two magnons and the photon are $g_{ic}=g_i\mu_B\sqrt{\mu_0\omega_cS_iN/2\hbar V}(i=A, B)$. $N$ and $V$ are the number of spins on each sublattice and the volume of the cavity, respectively. The detailed quantization of the Hamiltonian in Eq.~(\ref{H}) is given in Sec.~A of the supplementary material.

We can easily obtain the quantum Langevin equations (QLEs) from the Hamiltonian~(\ref{H_bosonic}) when taking into account the dissipation-fluctuation processes,
\begin{align}
&\frac{da}{dt}=-(\kappa_a + i\omega_a)a - ig_{ab} b^\dagger -ig_{ac} c^\dagger +\sqrt{2\kappa_a}a^{\mathrm{in}},\nonumber\\
&\frac{db}{dt}=-(\kappa_b + i\omega_b)b - ig_{ab} a^\dagger -ig_{bc} c +\sqrt{2\kappa_b}b^{\mathrm{in}},\nonumber\\
&\frac{dc}{dt}=-(\kappa_c + i\omega_c)c - ig_{ac} a^\dagger -ig_{bc} b +\sqrt{2\kappa_c}c^{\mathrm{in}},
\label{lang}
\end{align}
where $\kappa_a,~\kappa_b,~\kappa_c$ are the dissipation rates of magnon modes $a,~b$ and the photon mode $c$, respectively. The input quantum noises $a^{\mathrm{in}},~b^{\mathrm{in}},~c^{\mathrm{in}}$ arise from the coupling with surrounding environments. Here we assume three modes are in ordinary zero temperature such that the quantum noise can be characterized by the following correlation function $\langle v^{in}(t)v^{in\dagger}(t')\rangle=\delta(t-t')~(v=a,b,c)$ with zero mean average.

The Gaussian property of input quantum noises and linearized dynamics guarantee the steady state of the system is a Gaussian state which can be fully characterized by a canonically conjugate pair of variables for each mode, the so-called $X$ and $P$ quadratures, in phase space. The QLEs describing the motion of quadratures can be written as
\begin{equation}
\dot{u}=Mu+\Gamma u^{\mathrm{in}},
\end{equation}
where $u=(X_a,Y_a,X_b,Y_b,X_c,Y_c)^{\mathrm{T}}$ (the superscript T represents the transpose) with quadrature operators defined by $X_v=(v+v^\dagger)/\sqrt{2},~ Y_v=(v-v^\dagger)/\sqrt{2}i$ ($v=a,b,c$). The corresponding quadrature operators of noise are $u=(X_a^{\mathrm{in}},Y_a^{\mathrm{in}},X_b^{\mathrm{in}},Y_b^{\mathrm{in}},X_c^{\mathrm{in}},Y_c^{\mathrm{in}})^{\mathrm{T}}$ with $X_v^{\mathrm{in}}=(v^{\mathrm{in}}+v^{\mathrm{in}\dagger})/\sqrt{2},~ Y_v^{\mathrm{in}}=(v^{\mathrm{in}}-v^{\mathrm{in}\dagger})/\sqrt{2}i$ ($v=a,b,c$). The diagonal dissipative matrix $\Gamma=\mathrm{diag}(\sqrt{2\kappa_a},\sqrt{2\kappa_a},\sqrt{2\kappa_b},\sqrt{2\kappa_b},\sqrt{2\kappa_c},\sqrt{2\kappa_c})$ and the drift matrix is given by
\begin{equation}
M=\left (
\begin{array}{cccccc}
  -\kappa_a & \omega_a & 0 & -g_{\mathrm{ab}} & 0 & -g_{\mathrm{ac}} \\
  -\omega_a & -\kappa_a & -g_{\mathrm{ab}} & 0 & -g_{\mathrm{ac}} & 0 \\
  0 & -g_{\mathrm{ab}} & -\kappa_b & \omega_b & 0 & g_{\mathrm{bc}} \\
  -g_{\mathrm{ab}} & 0 & -\omega_b & -\kappa_b & -g_{\mathrm{bc}} & 0 \\
  0 & -g_{\mathrm{ac}} & 0 & g_{\mathrm{bc}} & -\kappa_c & \omega_c \\
  -g_{\mathrm{ac}} & 0 & -g_{\mathrm{bc}} & 0 & -\omega_c & -\kappa_c
\end{array} \right ).
\label{Mxyz}
\end{equation}
The system is stable only if all the eigenvalues of drift matrix $M$ have negative real parts. The general stability condition can be derived from the Routh-Hurwitz criterion~\cite{Dejesus1987}. The chosen parameters in the present work satisfy the stability condition. The steady state of the system is a Gaussian state with zero mean average, which can be entirely characterized by a $6\times6$ covariance matrix (CM) $V$ with components $V_{ij}=\langle u_i (\infty)u_j (\infty)+ u_j(\infty) u_i (\infty)\rangle /2$ ($i,j=1,2,\ldots,6$). The steady-state CM can be obtained by solving the Lyapunov equation $MV+VM^{\mathrm{T}}=-D$, where $D=\mathrm{diag}(\kappa_a,\kappa_a,\kappa_b,\kappa_b,\kappa_c,\kappa_c)$ is diffusion matrix charactering the stationary noise correlations defined by $\langle v_i(t)v_j(t')+v_j(t')v_i(t)\rangle/2=D_{ij}\delta(t-t')$ ($v=\Gamma u^{in}$)~\cite{Vitali2007}. 

We adopt the logarithmic negativity $E_N$ to quantify the bipartite entanglement, which has been proposed as a reliable quantitative estimate of continuous variable entanglement~\cite{Adesso2004}. The definition of $E_N$ is given by  $E_N=\max \{ 0, -\ln 2\eta^- \}$, where $\eta^-= \sqrt{ \sum (V') - \left [ \sum (V')^2-4\det V' \right ]^{1/2}}/\sqrt{2}$ is the smallest symplectic eigenvalue of the partially transposed CM, with $\Sigma(V')\equiv\det{V_a}+\det{V_b}-2\det{V_{ab}}$. Here we have considered the reduced CM of two modes of interest $V'=\left(
   \begin{array}{cc}
   V_a & V_{ab}\\
   V_{ab}^\mathrm{T} & V_b
   \end{array}
   \right)$, where $V_a$ and $V_b$ are $2\times 2$ block matrices corresponding to the reduced states of modes $a$ and $b$, respectively. To measure Gaussian steering, we employ the intuitive and computable quantification recently put forward in Ref.~\cite{Kogias2015}, which is a necessary and sufficient criterion for arbitrary bipartite Gaussian states under Gaussian measurements. The steering in two directions is given by $\mathcal{G}^{a\rightarrow b}=\max{\{0,S(2V_a)-S(2V')\}},~\mathcal{G}^{b\rightarrow a}=\max{\{0,S(2V_b)-S(2V')\}}$, with $S(\sigma)=\frac{1}{2}\ln{\det{\sigma}}$ being the R\'enyi-2 entropy~\cite{RenyiEntropy}. $\mathcal{G}^{a \rightarrow b}>0$ ($\mathcal{G}^{b \rightarrow a}>0$) demonstrates that the bipartite Gaussian state characterized by the covariance matrix $V'$ is steerable from mode $a$ ($b$) to mode $b$ ($a$) by Gaussian measurements on mode $a$ ($b$), and the larger value of $\mathcal{G}$ implies the stronger Gaussian steerability.

\section{3 Results}\label{RESULTS}

In a ferrimagnet the spin amplitudes on the two sublattices are different, i.e., $S_A\neq S_B$. This results in unequal coupling strengths of the spin waves with the photon mode  ($g_{ac}\neq g_{bc}$). This is in contrast to an antiferromagnet in which the spin amplitudes are equal. Thus, antiferromagnet can be regarded as a special type of ferrimagnet characterized by equal saturation magnetizations of both sublattices.
\begin{figure}
\includegraphics[width=0.8\columnwidth]{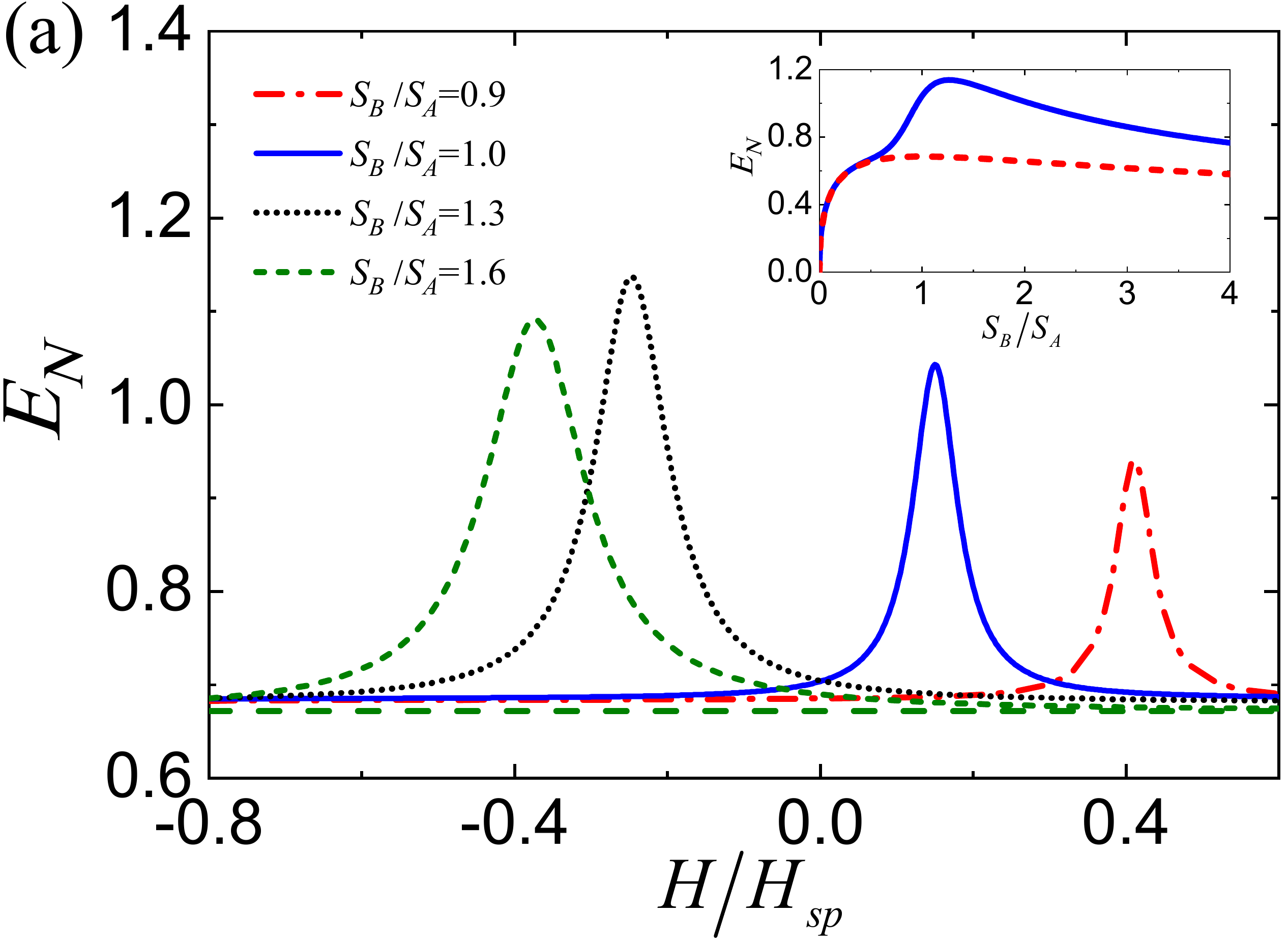}
\includegraphics[width=0.8\columnwidth]{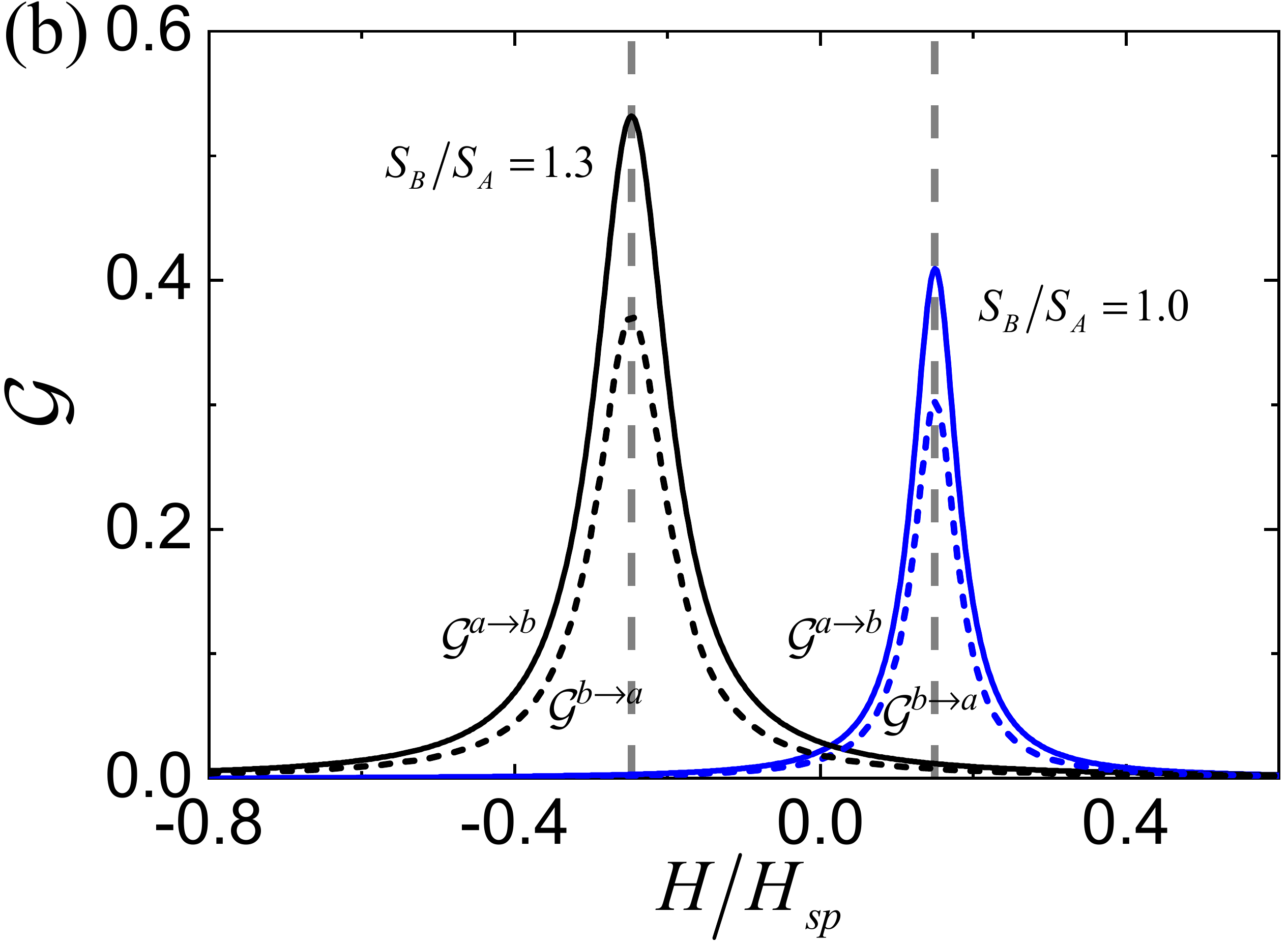}
\caption{(a) Variation of the magnon-magnon entanglement measure $E_N$ with $H/H_{sp}$ for the magnons coupled to the photon mode $c$ and for different spin amplitudes $S_B/S_A$. The green long dashed horizontal line indicates the degree of entanglement between the magnons in the absence of the mode $c$ for $S_B/S_A=1.6$. The inset shows the variation of the maximum value of the magnon-magnon entanglement with the ratio $S_B/S_A$ when optimizing the external field. The blue solid and red dashed traces correspond to the cases when the magnons are respectively coupled to and decoupled from the photon mode. (b) Variation of the steering parameters $\mathcal{G}^{a\rightarrow b}$ (solid lines) and $\mathcal{G}^{b \rightarrow a}$ (dashed lines) with $H/H_{sp}$ in the presence of the photon mode $c$ for $S_B/S_A=1.3$ and $S_B/S_A=1$. Other parameters are: photon frequency $\omega_c/H_{sp} = 0.85$,~$\omega_{a,\mathrm{an}}=0.0163\omega_{b,\mathrm{ex}},~\kappa_a=\kappa_b=0.001\omega_{b,\mathrm{ex}},~\kappa_c=0.003\omega_{b,\mathrm{ex}},~g_{ac}=0.01\omega_{b,\mathrm{ex}}$, thus $~\omega_{b,\mathrm{an}}=(S_B/S_A)\omega_{a,\mathrm{an}},~g_{bc}=g_{ac}\sqrt{S_B/S_A}$, and $g_{ab}=\sqrt{S_B/S_A}\omega_{b,\mathrm{ex}}$. 
We take $\omega_{b,\mathrm{ex}}=1$ throughout, so all parameters are measured relative to this quantity.}
\label{Fig2}
\end{figure}

Figure~\ref{Fig2}(a) shows the dependence of the stationary entanglement $E_N$ between two magnon modes $a$ and $b$ on the external field $H/H_{sp}$ for different ratios of the spin amplitudes $S_B/S_A$. Here, $H_{sp}=\sqrt{\omega_{\mathrm{b,an}}(\omega_{\mathrm{b,an}}+2\omega_{b,\mathrm{ex}})}$ denotes the spin-flop field in the AFM, i.e., $S_B=S_A$. In the absence of the photon mode $c$, some degree of the magnon-magnon entanglement is obtained $(E_{N}=0.68)$, which is independent of $H/H_{sp}$. When the two-sublattice ferrimagnet is coupled to the photon mode the entanglement is enhanced relative to the value obtained in the absence of the photon mode. The enhancement is seen to occur at a specific value of $H/H_{sp}$ that $E_{N}$ displays peaks emerging from the background value of $E_{N}=0.68$. The position of the peaks varies with $S_B/S_A$ that the peaks move from positive to negative values of $H/H_{sp}$ when $S_B/S_A$ is varied from $S_B/S_A<1$ to $S_B/S_A>1$. Moreover, the significant enhancement of entanglement occurs for $S_B/S_A>1/2$. This is shown in the inset of Fig.~\ref{Fig2}(a), where we plot the variation of the maxima of the $E_{N}$ peaks with $S_B/S_A$. It is evident that in the presence of the photon mode the maxima of $E_{N}$ (blue solid line) are significantly higher than those in the absence of the photon mode, and the enhancement clearly occurs for $S_B/S_A>1/2$.  The explanation will follow from the analysis of the populations of the Bogoliubov modes presented in Fig.~\ref{Fig5}.

The enhancement of entanglement results in the creation of quantum steering between the magnon modes. This is illustrated in Fig.~\ref{Fig2}(b) which shows the Gaussian steering parameters $\mathcal{G}^{a\rightarrow b}$ and $\mathcal{G}^{b \rightarrow a}$ as a function of $H/H_{sp}$ for $\kappa_{a}=\kappa_{b}$ and two different values of $S_B/S_A$; $S_B/S_A=1.3$ and $S_B/S_A=1$. It is seen that in the absence of the photon mode quantum steering between the magnon modes cannot be created if the modes are damped with the same rates, $\kappa_{1}=\kappa_{2}$. The effect of coupling the magnon modes to the photon mode is seen to create quantum steering between the modes. In the case of ferrimagnet $(S_B/S_A=1.3)$ quantum steering is stronger than that for antiferromagnet $(S_B/S_A=1)$. However, in both cases, an asymmetric two-way steering is observed that  
$\mathcal{G}^{a\rightarrow b}$ is always larger than $\mathcal{G}^{b \rightarrow a}$. Note that this system creates asymmetric quantum steering without imposing additional conditions of asymmetric losses or noises of the subsystems. In addition, we can also see that the enhancement of steering is maximized at the same value of $S_B/S_A$ at which the entanglement is maximized.  
\begin{figure}
\includegraphics[width=0.83\columnwidth]{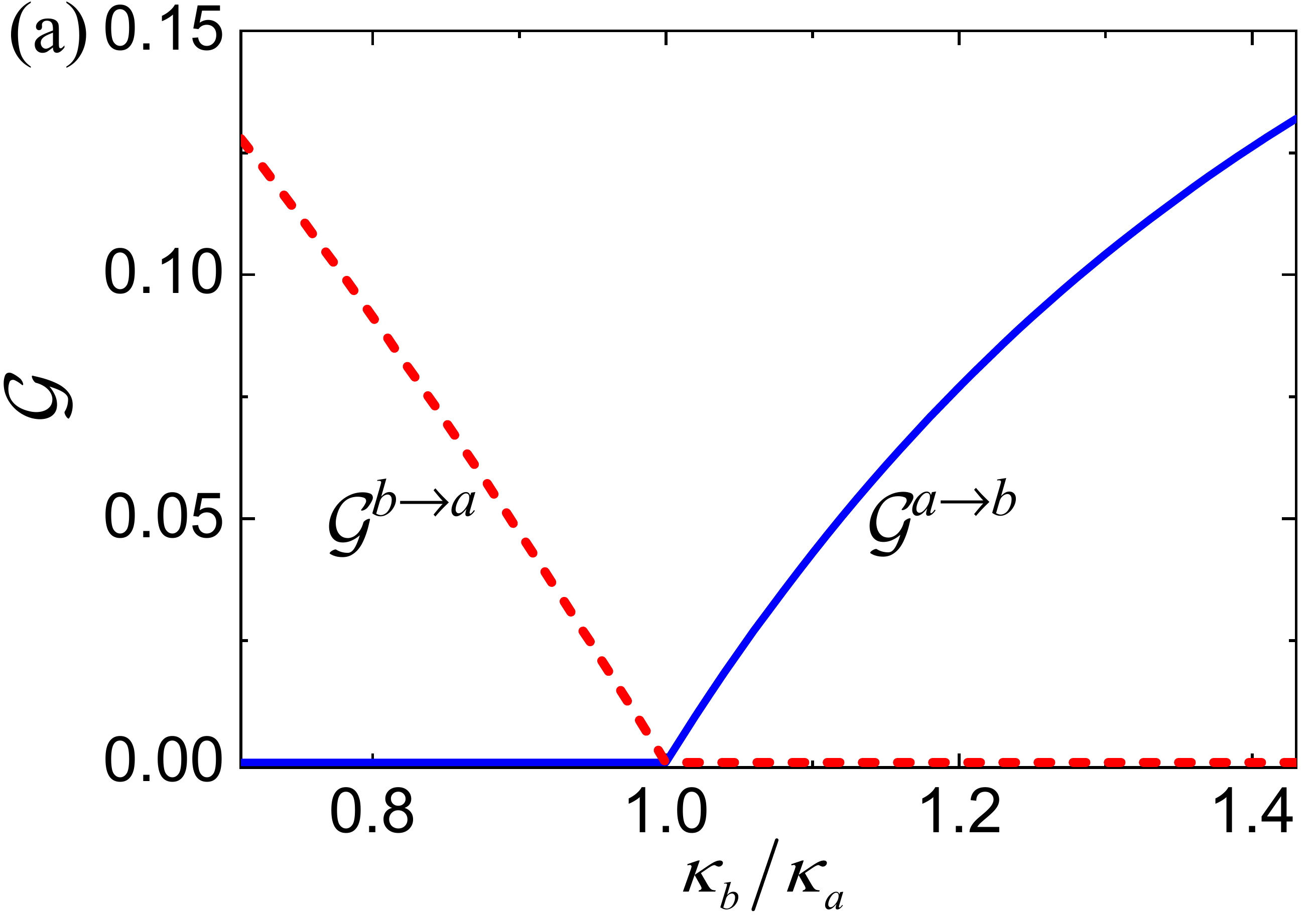}
\includegraphics[width=0.83\columnwidth]{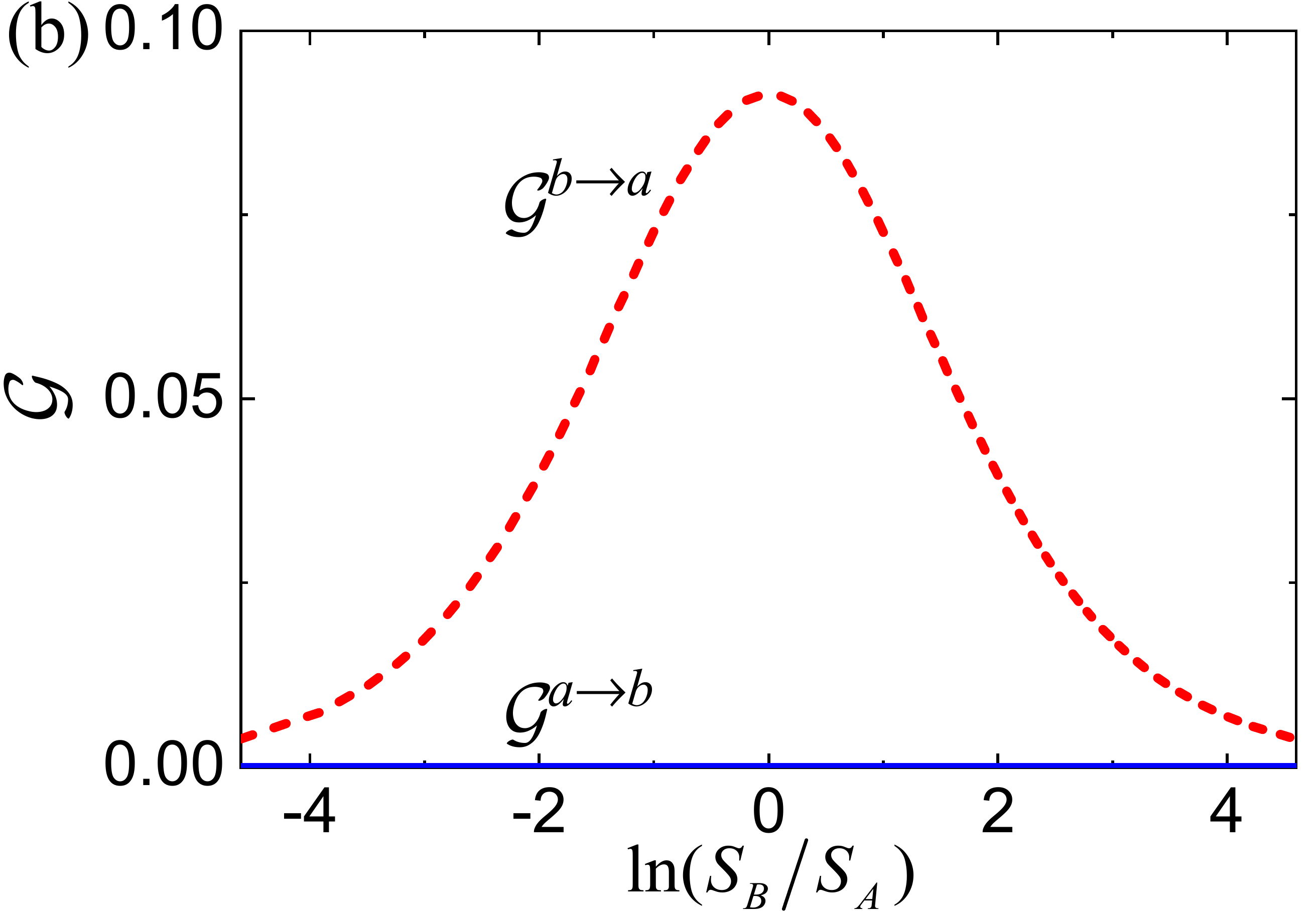}
\caption{(a) Variation of $\mathcal{G}^{a\rightarrow b}$ and $\mathcal{G}^{b \rightarrow a}$ with the ratio $\kappa_{b}/\kappa_{a}$ when the magnon modes are decoupled from the photon mode $c$ and for $S_B/S_A=1$. (b) Illustration of the variation of the one-way steering parameter $\mathcal{G}^{b \rightarrow a}$ with $\ln{(S_B/S_A)}$ for $\kappa_b/\kappa_a=0.8$. Other parameters are:  $\omega_{a,\mathrm{an}}=0.0163\omega_{b,\mathrm{ex}},~\kappa_a=0.001\omega_{b,\mathrm{ex}},~\omega_{b,\mathrm{an}}=(S_B/S_A)\omega_{a,\mathrm{an}}$, and $g_{ab}=\sqrt{S_B/S_A}\omega_{b,\mathrm{ex}}$.}
\label{Fig3}
\end{figure}

We have seen that in the absence of the photon mode there is no quantum steering between magnons when the modes are damped with the same rates. However, if the modes are damped with different rates $(\kappa_{a}\neq\kappa_{b})$, one-way steering with either $\mathcal{G}^{a\rightarrow b}>0,\ \mathcal{G}^{b\rightarrow a}=0$ or $\mathcal{G}^{b\rightarrow a}>0,\ \mathcal{G}^{a\rightarrow b}=0$ can be created in the absence of the photon mode. This is shown in Fig.~\ref{Fig3} where in frame (a) we plot $\mathcal{G}^{a\rightarrow b}$ and $\mathcal{G}^{b\rightarrow a}$ as a function of $\kappa_{b}/\kappa_{a}$ for $S_B/S_A=1$, and in frame (b) we plot $\mathcal{G}^{a\rightarrow b}$ and $\mathcal{G}^{b\rightarrow a}$ as a function of $S_B/S_A$ for $\kappa_{a}\neq \kappa_{b}$. The effect of unequal damping rates is seen to create one-way steering between the magnon modes. Once $\kappa_{a}\neq \kappa_{b}$, one-way steering is created with the maximum value attained at $S_A=S_B$.
 The role of dissipation in spintronics and magnetism is not well understood as it is of fundamental interest to determine the efficiency of spintronic devices. For a ferromagnet, the damping can be well characterized by ferromagnetic resonance. However, for a ferrimagnet/antiferromagnet the magnetic resonance can only give limited information about the relaxation of collective excitation and fails to account for the damping of each sublattice. We suggest that observation of an asymmetric EPR steering can be used as an alternative method of determining damping rates of the two sublattices. This may serve as a useful supplement to the traditional methods of detection of magnetic damping and could further help us to understand the inter-sublattice spin pumping effect among the sublattices of a magnet.

\section{4 Discussion}\label{Discussion}

In order to understand the essential physics of the results presented in Figs.~\ref{Fig2} and \ref{Fig3}, in particular, the process of enhancement of the magnon-magnon entanglement, the dependence of the positions of the peaks on $H/H_{sp}$, and the creation of asymmetric steering between the magnons, we diagonalize the Hamiltonian (\ref{H_bosonic}) by introducing two Bogoliubov modes
\begin{eqnarray}
\alpha &=& \cosh r a + \sinh r b^\dagger ,\nonumber\\
\beta &=& \sinh r a^\dagger + \cosh r b ,
\end{eqnarray}
where $\tanh 2 r=2g_{ab}/(\omega_a + \omega_b)$, and $\alpha$ and $\beta$ satisfy the bosonic commutation relations. In terms of the Bogoliubov modes the Hamiltonian (\ref{H_bosonic}) becomes
\begin{eqnarray}
\mathcal{H} &=& \omega_\alpha \alpha^\dagger \alpha  + \omega_\beta \beta^\dagger \beta + \omega_c c^\dagger c \nonumber\\
&+& g_{\alpha c} \left(\alpha c+\alpha^\dagger c^\dagger\right) +g_{\beta c}\left( \beta^\dagger c+\beta c^\dagger\right) ,\label{eq}
\end{eqnarray}
where $g_{\alpha c}=g_{ac}\cosh r-g_{bc}\sinh r,~g_{\beta c}=-g_{ac}\sinh r+g_{bc}\cosh r$ are the coupling strengths between the Bogoliubov modes and the photon mode,
and $\omega_\alpha=(\omega_a-\omega_b+\sqrt{(\omega_a+\omega_b)^2-4g_{ab}^2})/2$,~$\omega_\beta=(-\omega_a+\omega_b+\sqrt{(\omega_a+\omega_b)^2-4g_{ab}^2})/2$ represent respectively the optical (higher frequency) and acoustic (lower frequency) magnon bands.

\begin{figure}
\includegraphics[width=1.05\columnwidth]{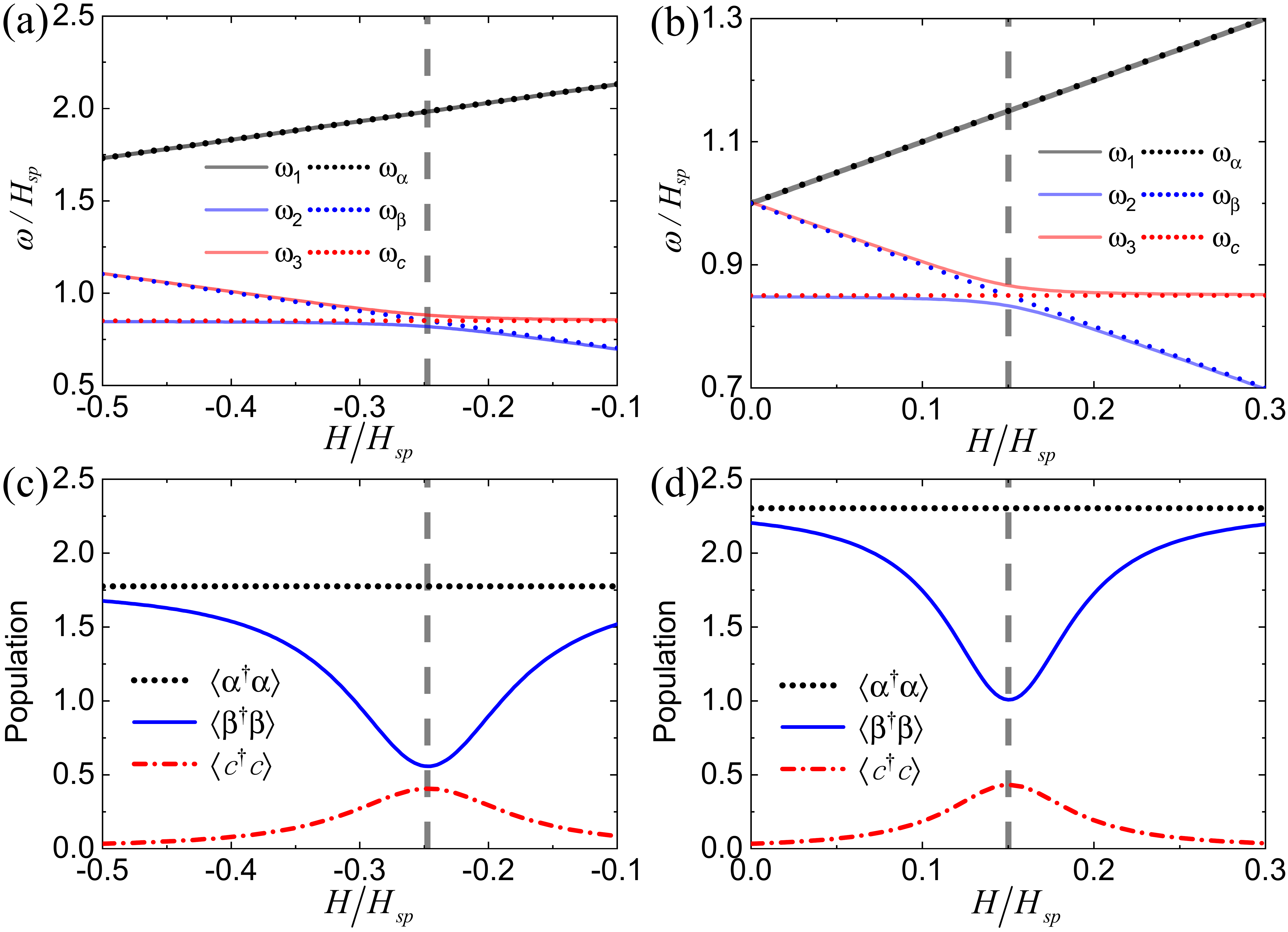}
\caption{(a)-(b) Variation of the frequencies of the Bogoliubov modes $\omega_{\alpha,\beta}$, the photon mode $\omega_c$, and the eigenmodes of the coupled system $\omega_{1,2,3}$ with $H/H_{sp}$ for two different values of the ratio $S_B/S_A$: (a) $S_B/S_A=1.3$ and (b) $S_B/S_A=1.0$. Note that $\omega_\beta$ and $\omega_c$ superpose to form the $\omega_{2}$ and $\omega_3$, while $\omega_\alpha$ is left unchanged and forms $\omega_1$. (c)-(d) The populations of modes $\alpha,\beta$ and $c$ plotted as a function of $H/H_{sp}$ for two different values of the ratio $S_B/S_A$: (c) $S_B/S_A=1.3$ and (d) $S_B/S_A=1$. The populations refer to the mean numbers of magnons and photons, i.e., $\langle \alpha^\dagger \alpha\rangle,~\langle \beta^\dagger \beta\rangle,~\langle c^\dagger c\rangle$, which are dimensionless quantities. Other parameters are the same as in Fig.~\ref{Fig2}.}
\label{Fig4}
\end{figure}

In Fig.~\ref{Fig4} we plot the eigenfrequencies $\omega_{1,2,3}$ of the coupled system together with the frequencies of the Bogoliubov modes $\omega_{\alpha,\beta}$ and the photon mode $\omega_c$ as a function of $H/H_{sp}$ for the case $S_B/S_A=1.3$ (frame (a)), comparing them with the case of antiferromagnet, $S_B/S_A=1$ (frame (b)). We also plot, in frames (c) and (d), the populations of the Bogoliubov modes and the photon mode. Referring to the results for the magnon-magnon entanglement and quantum steering (Fig.~\ref{Fig2}), we see that the enhancement of the entanglement and steering occurs when the acoustic magnon mode $\beta$ is resonant with the photon mode $c$, i.e., when $\omega_\beta=\omega_c$ which takes place at $H/H_{sp}\approx -0.25, ~0.15$ for $S_B/S_A=1.3, ~1.0$, respectively. Note that the magnon states should be around the ground state when using Holstein-Primakoff transformation, and therefore, the external static field $H$ is negative to guarantee the stability of the ground state for the case when $S_B/S_A>1$.

\subsection{4.1 Role of the cavity cooling}\label{cavity cooling}

The enhancement of the magnon-magnon entanglement can be well understood through cavity cooling of the Bogoliubov modes toward their joint ground state~\cite{Yuan2020PRB}. We can reformulate the magnon eigenmodes $\alpha=S(r)aS^\dagger(r),~\beta=S(r)bS^\dagger(r)$ where $S(r)={\rm exp}[r(ab-a^\dagger b^\dagger)]$ is a unitary two-mode squeezing operator. Clearly, the superposition magnon modes $\alpha,~\beta$ result from a two-mode squeezing transformation of modes $a$ and $b$ with squeezing parameter $r$. For an ideal two-mode squeezed vacuum state $|r\rangle=S(r)|0_a,0_b\rangle$, the entanglement is known as $E_N=2r={\rm arctanh}[2g_{ab}/(\omega_a + \omega_b)]$. We can see that the ratio of $g_{ab}/\omega_{a,b}$ determines the degree of squeezing as well as entanglement between modes $a$ and $b$. With our parameters $g_{ab}=\sqrt{S_B/S_A}\omega_{b,\mathrm{ex}},~\omega_a + \omega_b=(S_B/S_A+1)(\omega_{b,\mathrm{ex}}+\omega_{a,\mathrm{an}})$, the maximal entanglement (ground-state) should be $E_N ={\rm arctanh}[2g_{ab}/(\omega_a + \omega_b)]\approx1.91$ for $S_B/S_A=1.6$. However, in the absence of light the achieved state is not an ideal two-mode squeezed vacuum state, we can calculate the logarithmic negativity based on its CM matrix and obtain $E_N\approx 0.67$ for the case $S_B/S_A=1.6$, as indicated by the green long dashed horizontal line in Fig.~\ref{Fig2}(a). When the magnons couple to the photon mode, we can see from Eq.~(\ref{eq}) that the photons play as a cold bath to cool the mode $\beta$ toward its ground state (i.e., the squeezed vacuum state of modes $a,~b$) by the beam-splitter-type coupling $g_{\beta c}( \beta^\dagger c+\beta c^\dagger)$, and consequently the entanglement of the state is enhanced.

To further analyze the cooling effect, we may refer to the populations of the Bogoliubov modes $\alpha,~\beta$ and the photon mode $c$, shown in frames $(c)$ and $(d)$ of Fig.~\ref{Fig4}. We see that the population of the mode $\beta$ is significantly reduced and reaches the minimum value at the anticrossing point. At this point the population of the photon mode $\langle c^\dagger c\rangle$ takes the maximum value. This means that the cooling effect is most efficient at the anticrossing point due to the strong coupling between the magnon modes and the photon mode. As a result, the maximal enhancement of entanglement is obtained at the same point. It is worth pointing out that in the case of the ferrimagnet $(S_B/S_A\neq 1)$, shown in Fig.~\ref{Fig4}(c), the population $\langle \beta^\dagger \beta\rangle$ achieves minimum value which is smaller than that achieved in antiferromagnet $(S_B/S_A=1)$, shown in Fig.~\ref{Fig4}(d). This explains why the enhancement of entanglement and quantum steering for the ferrimagnetic case illustrated in Fig.~\ref{Fig2} is much stronger than those for the antiferromagnetic. Note that the population of the mode $\alpha$ remains almost unchanged since the frequency of the mode $\omega_\alpha=\omega_1$ which is decoupled from the frequency of the photon mode.
\begin{figure}[b]
\includegraphics[width=0.75\columnwidth]{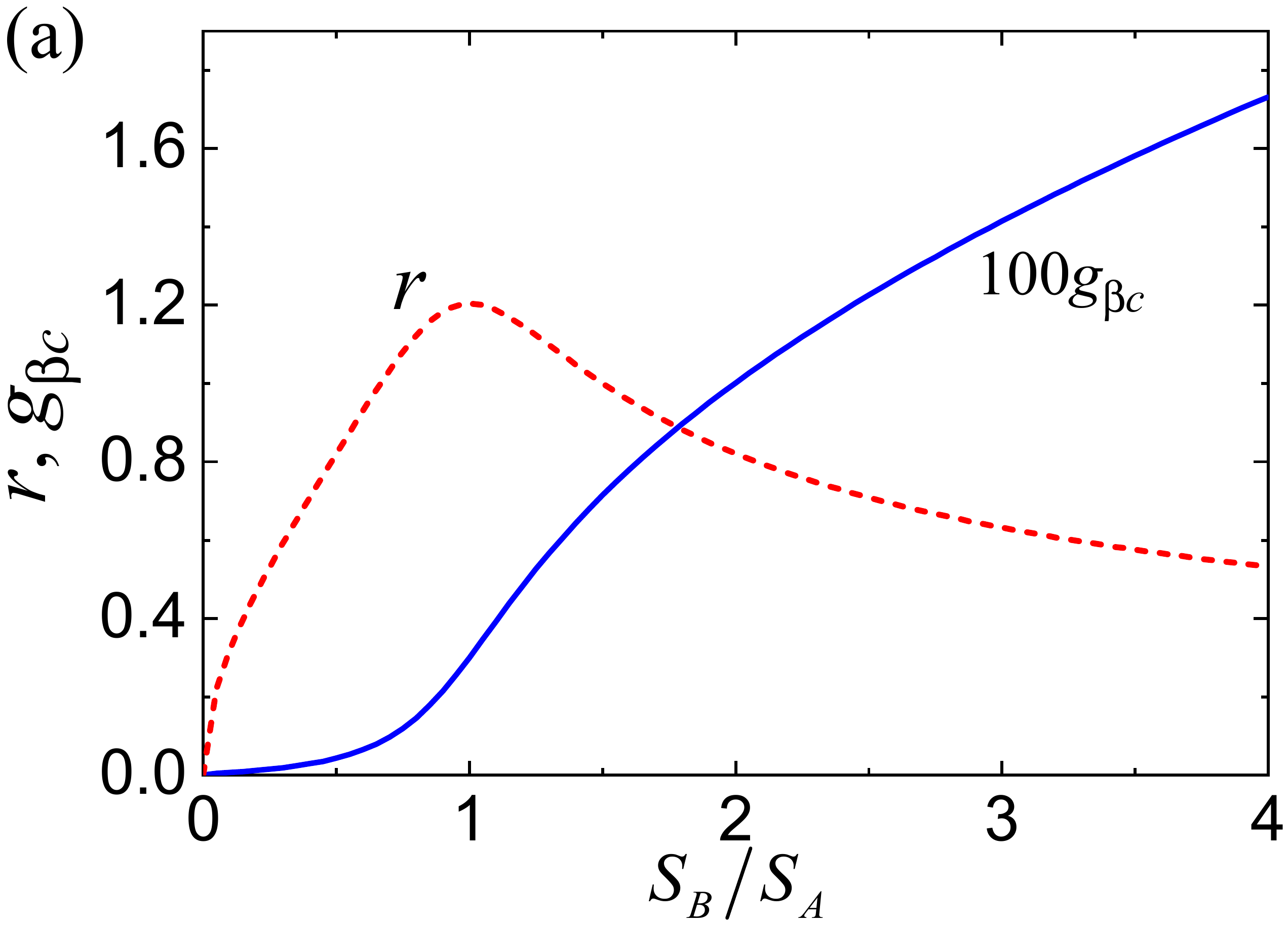}
\includegraphics[width=0.75\columnwidth]{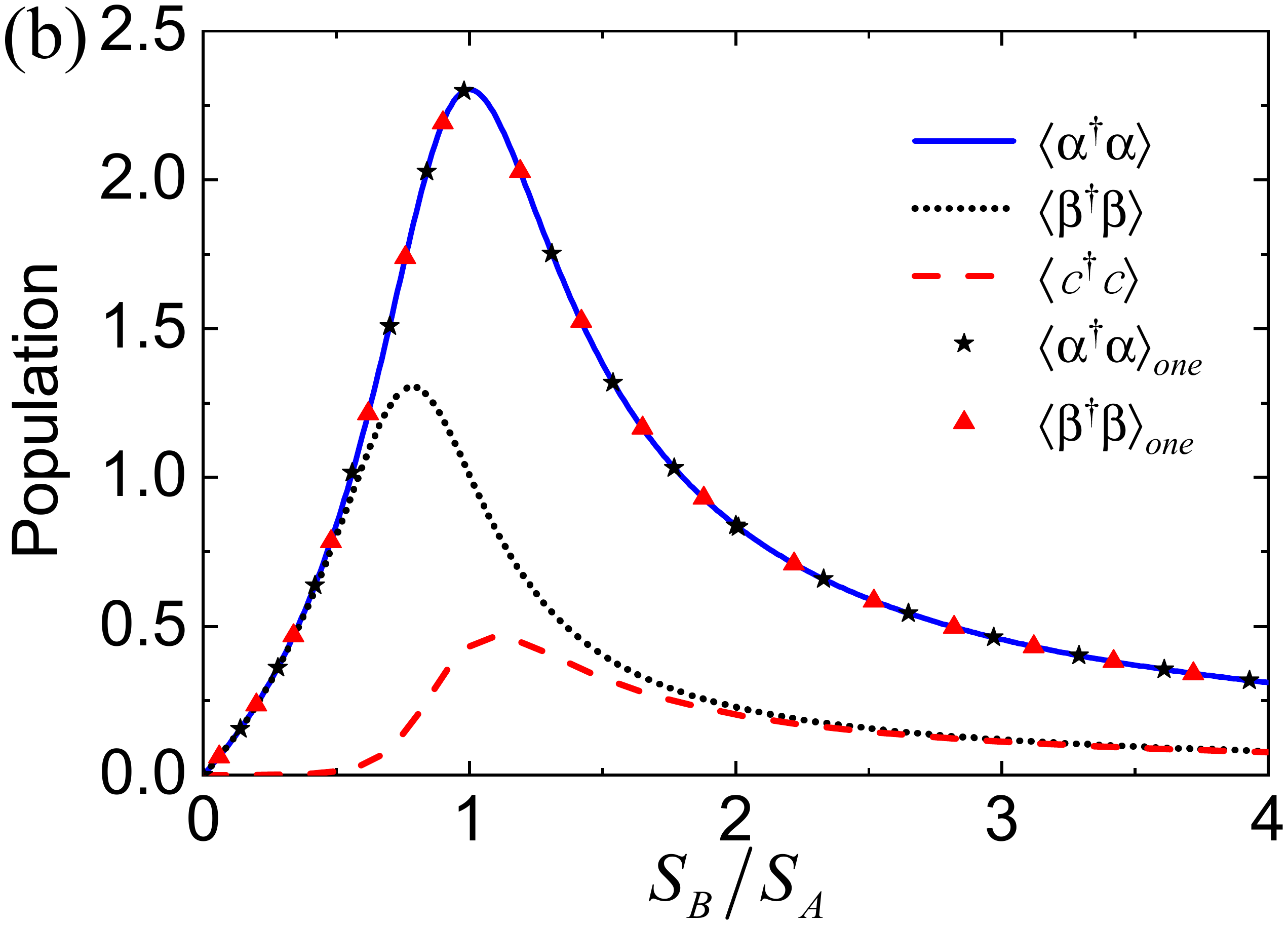}
\caption{(a) The squeezing parameter $r$ and the coupling strength $g_{\beta c}$ plotted as a function of spin amplitudes $S_B/S_A$ for the optimized external field $H$. (b) The populations of the modes $\alpha,\beta$ and $c$ when the magnon modes are coupled to the photon mode. Black stars represent population of the mode $\alpha$  $(\langle \alpha^\dagger \alpha\rangle_{one})$ and red triangles represent population of the mode $\beta$ $(\langle \beta^\dagger \beta\rangle_{one})$ when the magnon modes are decoupled from the photon mode. The parameters are the same as  in Fig.~\ref{Fig2}.}
\label{Fig5}
\end{figure}

The fact that among three modes only the mode $\beta$ is cooled in the system also helps us to understand why the magnon-magnon entanglement is almost equal to that in the absence of light for $S_B/S_A<1/2$, as seen in the inset of Fig.~\ref{Fig2}(a). Figure~\ref{Fig5} illustrates the variation of the squeezing parameter $r$ and the coupling strength $g_{\beta c}$ with $S_B/S_A$. It is apparent that the squeezing parameter $r$ and especially the coupling strength $g_{\beta c}$ are small at $S_B/S_A<1/2$. Moreover, in this region the population of the photon mode $c$ is nearly zero. Also in this region the populations of the modes $\alpha$ and $\beta$ are almost the same and independent of whether the modes are coupled to the photon mode or not, i.e., $\langle \alpha^\dagger \alpha\rangle\approx\langle \alpha^\dagger \alpha\rangle_{one}$ and $\langle \beta^\dagger \beta\rangle \approx\langle \beta^\dagger \beta\rangle_{one}$. Clearly the cooling effect is completely suppressed in this region. Therefore, no enhancement of entanglement is observed for $S_B/S_A<1/2$ even in the presence of the photon mode.  

The behavior of the populations of the modes $\alpha$ and $\beta$, shown in Figs.~\ref{Fig4}(c) and (d), can also help us to explain why the steering $\mathcal{G}^{a\rightarrow b}$ is always larger than $\mathcal{G}^{b\rightarrow a}$, the property shown in Fig.~\ref{Fig2}(b). Since the difference of populations of modes $a$ and $b$ can be expressed in terms of the populations of the superposition modes $\alpha$ and $\beta$, i.e., $\langle a^\dagger a\rangle -\langle b^\dagger b\rangle=\langle \alpha^\dagger \alpha\rangle -\langle \beta^\dagger \beta\rangle$, the fact that the mode $\beta$ is cooled by the photons while the mode $\alpha$ is decoupled from photons leads to the population  $\langle \alpha^\dagger \alpha\rangle$ always larger than $\langle \beta^\dagger \beta\rangle$. This means that $\langle a^\dagger a\rangle>\langle b^\dagger b\rangle$, and thus $\mathcal{G}^{a\rightarrow b}>\mathcal{G}^{b\rightarrow a}$. More detailed analysis is given in Sec.~B of the supplementary material. Note that in the absence of light, steering between two magnons $\mathcal{G}^{a\rightarrow b}=\mathcal{G}^{b\rightarrow a}=0$, as we assume the identical dissipation rates of two magnons $\kappa_a=\kappa_b$ (see Sec.~B of the supplementary material for the analytical expressions). Thus, instead of the enhancement of entanglement, steering between two magnons that have equal dissipation rates can be only created with the assistance of photons. This provides an alternative way to create asymmetric steering beyond adding asymmetric losses or noises to the subsystems.

\subsection{4.2 Role of the first-order coherence between the magnon and photon modes}\label{photon mode}

We have seen that in the presence of the photon mode the entanglement and quantum steering are much more enhanced in the case of ferrimagnet $(S_{B}/S_{A}\neq 1)$ than for antiferromagnet $(S_{B}/S_{A}= 1)$. We may attribute the difference in the enhancement of entanglement to the presence of the first-order coherence between the magnon mode $b$ and the photon mode $c$. The degree of the first-order coherence between modes $b$ and $c$ is given by
\begin{equation}
 \gamma_{b,c}^{(1)}= \frac{|\langle b^\dagger c\rangle|}{\langle b^\dagger b\rangle^{1/2} \langle c^\dagger c\rangle^{1/2}} .
 \end{equation} 

\begin{figure}[h]
\includegraphics[width=1.05\columnwidth]{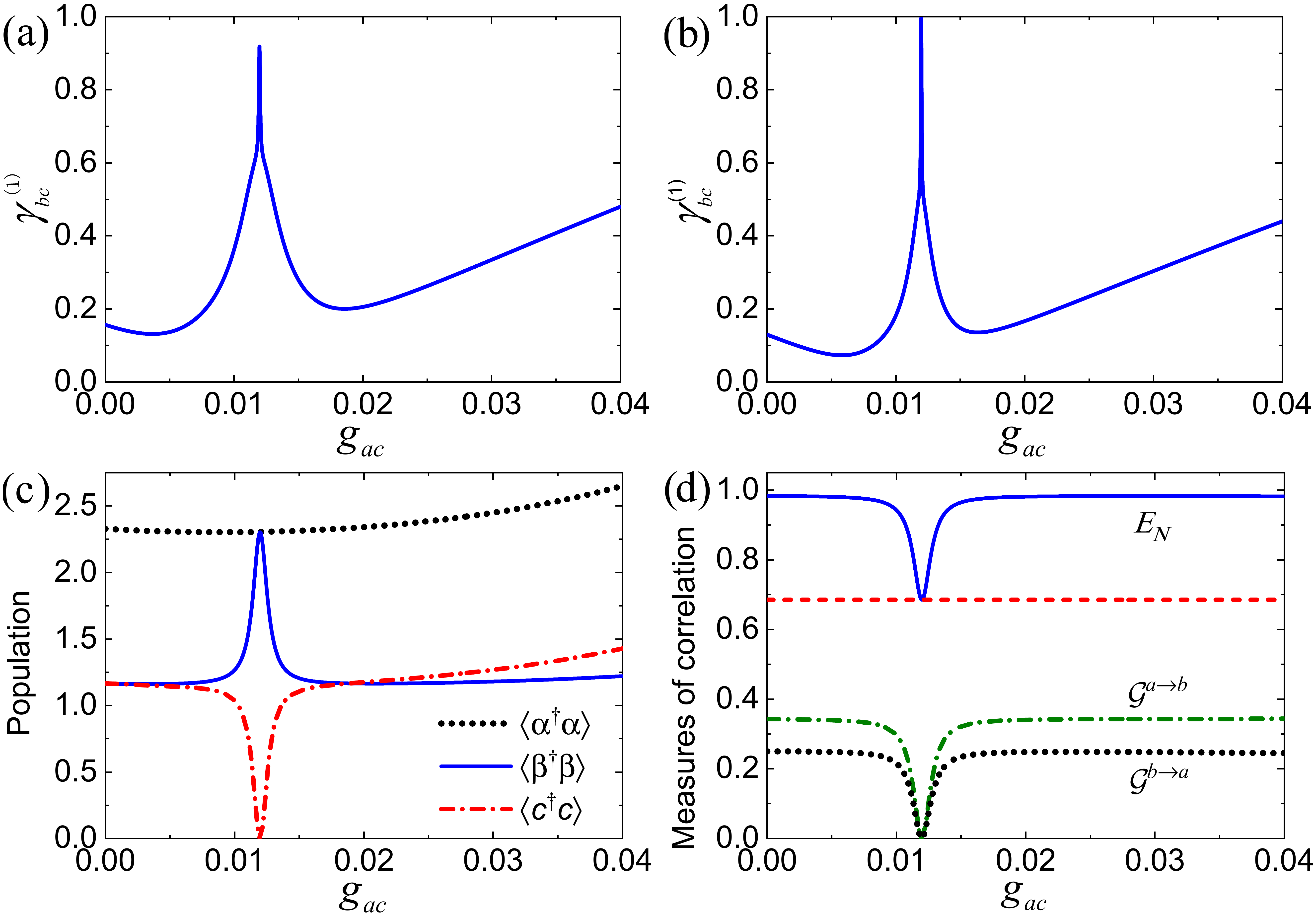}
\caption{(a) Degree of the first-order coherence $\gamma_{bc}^{(1)}$ between the magnon mode $b$ and the photon mode $c$ plotted as a function of $g_{ac}$ with damping rates $\kappa_a=\kappa_b=0.001\omega_{b,\mathrm{ex}}$, and $\kappa_c=0.003\omega_{b,\mathrm{ex}}$.  (b) same to (a) but with $\kappa_a=\kappa_b=\kappa_c=0.001\omega_{b,\mathrm{ex}}$. (c) Variation of the populations of the modes $\alpha,~\beta$ and $c$ with $g_{ac}$. (d) Variation of the magnon-magnon entanglement measure $E_N$ (blue solid) and the steering parameters $\mathcal{G}^{a\rightarrow b}$ (green dash-dotted) and $\mathcal{G}^{b\rightarrow a}$ (black dotted) with the coupling strength $g_{ac}$ when the magnon modes are coupled to the photon mode $c$. The horizontal red dashed line indicates the magnon-magnon entanglement when the magnons are decoupled from the mode $c$.  Other parameters are $\omega_{a,\mathrm{an}}=\omega_{b,\mathrm{an}}=\omega_{\mathrm{an}}=0.0163\omega_{b,\mathrm{ex}},~\omega_c/H_{sp}=0.85,~H/H_{sp}=0.15,~g_{bc}=0.01\omega_{b,\mathrm{ex}},~g_{ab}=\omega_{b,\mathrm{ex}}$. }
\label{Fig6}
\end{figure}

Figure~\ref{Fig6}(a) shows $\gamma_{b,c}^{(1)}$ as a function of $g_{ac}$ for fixed $g_{bc}$ and the same choice of the damping rates as in Fig.~\ref{Fig2}. The degree of coherence exhibits a pronounced very narrow peak located at $g_{ac}\approx g_{bc}$ with the maximum value close to one $(\gamma_{b,c}^{(1)}=0.92)$. In other words, the modes are mutually almost perfectly coherent. The degree of coherence is significantly less than one $(\gamma_{b,c}^{(1)}<1/2)$ for all values of $g_{ac}$ different than $g_{bc}$. The reason for the enhancement of the coherence can be attributed to interference between two channels through which the modes $b$ and $c$ are coupled. One of the channels is the indirect coupling through the mode $a$; $b - a - c$, and the other channel is the direct coupling $b-c$. For fixed $g_{ab}$, an asymmetry between $g_{ac}$ and $g_{bc}$, $g_{ac}\neq g_{bc}$ makes the probability amplitudes of the two channels to be different. Let $g_{ac}\psi_{1}$ describes the probability amplitude of the $b-a-c$ channel and $g_{bc}\psi_{2}$ describes the probability amplitude of the $b-c$ channel. Then the total probability of the $b-c$ transition is
\begin{align}
|\Psi_{bc}|^{2} &= |g_{ac}\psi_{1} +g_{bc}\psi_{2}|^{2} \nonumber\\
&= g_{ac}^{2}|\psi_{1}|^{2} + g^{2}_{bc}|\psi_{2}|^{2} + 2g_{ac}g_{bc}Re{(\psi^{\ast}_{1}\psi_{2})},
\end{align}
which for fixed $\psi_{1}$ and $\psi_{2}$ attains a maximum for $g_{ac}=g_{bc}$.

The reason that for $g_{ac}\approx g_{bc}$ the degree of coherence is smaller than one can be attributed to the fact that the modes are damped with different rates, $\kappa_{b}\neq \kappa_{c}$, i.e., the channels could be partially distinguished through the different damping rates of the modes. When the modes decay with the same rates $\kappa_{b}=\kappa_{c}$, the modes can be mutually perfectly coherent for $g_{ac}\approx g_{bc}$. This is shown in Fig.~\ref{Fig6}(b) where we plot $\gamma_{b,c}^{(1)}$ for the same parameters as in Fig.~\ref{Fig6}(a) except that the damping rates are now equal, $\kappa_{a}=\kappa_{b}=\kappa_{c}$.

It is clearly seen that the degree of coherence equals unity for the value of $g_{ac}\approx g_{bc}$. In other words, the modes $b$ and $c$ are mutually perfectly coherent. When two modes are perfectly coherent, then maximal constructive or destructive interference between these modes could occur leading to the cancellation of the population of the mode $c$. This is shown in Fig.~\ref{Fig6}(c) where we plot the populations of the modes as a function of $g_{ac}$ for the same parameters as in Fig.~\ref{Fig6}(b). It is clear that the population of the magnon mode $c$ is reduced to zero for the value of $g_{ac}\approx g_{bc}$. At the same point, the population of the magnon mode $\beta$ exhibits a strong peak. Clearly, the magnon modes are heated instead of being cooled when $g_{ac}\approx g_{bc}$. This observation agrees with the picture of cavity cooling described by the Hamiltonian (8), in which the coupling strength of $g_{\beta c}=-g_{ac}\sinh r+g_{bc}\cosh r $ determines the performance of cooling effect and $g_{\beta c}=0$ when $g_{ac}\approx g_{bc}$. It is interesting that $\gamma_{bc}^{(1)}=1$ even if the mode $c$ is completely depopulated. This surprising behavior is an example of an interesting phenomenon called induced coherence without induced emission~\cite{heuer2015induced}. 

The zero value of the population $\langle c^\dagger c\rangle$ at the dip indicates that the interference is perfectly destructive and the population is not effectively transferred from $b$ to $c$. Thus, the system behaves as it would not be coupled to the photon mode. This is reflected in the behaviours of entanglement and quantum steering. In Fig.~\ref{Fig6}(d) we plot $E_{N}$, $\mathcal{G}^{a\rightarrow b}$ and $\mathcal{G}^{b\rightarrow a}$ as a function $g_{ac}$ for the same values of the parameters as in Fig.~\ref{Fig6}(b). We see that  $E_{N}$ and $\mathcal{G}^{a\rightarrow b}, \mathcal{G}^{b\rightarrow a}$ display a dip for $g_{ac}\approx g_{bc}$ instead of being constant over the entire $g_{ac}$. The minimum values of the dips corresponds to the values of the entanglement and steering obtained in the absence of the photon mode, $g_{ac} =g_{bc}= 0$. In other words, the values of the entanglement and quantum steering of the modes $a$ and $b$ at the dips are reduced to those obtained in the absence of the mode $c$. 

Although the modes $b$ and $c$ are mutually perfectly coherent, the visibility of the interference pattern can be zero such that the modes can be completely distinguishable. The visibility $\mathcal{V}$ and distinguishability $\mathcal{D}$ are related through the expression of complementarity, $|\mathcal{V}|^{2} +|\mathcal{D}|^{2}\leq 1$. When $|\mathcal{V}| =1$ then $|\mathcal{D}|=0$ which means that the modes are indistinguishable. On the other hand, when $|\mathcal{V}|=0$ then $|\mathcal{D}|=1$ that the modes are perfectly distinguished. The visibility and distinguishability are given by
\begin{align}
|\mathcal{V}| &= \frac{2|\langle b^{\dagger}c\rangle|} {\langle b^{\dagger}b\rangle + \langle c^{\dagger}c\rangle} ,\nonumber\\
|\mathcal{D}| &= \frac{|\langle b^{\dagger}b\rangle - \langle c^{\dagger}c\rangle|} {\langle b^{\dagger}b\rangle + \langle c^{\dagger}c\rangle} .\label{D}
\end{align}
Since for $g_{ac}\approx g_{bc}$ the population of the mode $c$ is zero, we see immediately that $|\mathcal{D}| =1$. In this case the modes are completely distinguishable. Correspondingly, no enhancement of entanglement and the creation of quantum steering could be observed. On the other hand, when $g_{ac}\neq g_{bc}$, the population $\langle c^{\dagger}c\rangle\neq 0$, indicating that the modes are no longer completely distinguishable. The lack of perfect distinguishability of the modes results in the enhancement of entanglement between the magnon modes.

Finally, we comment about the possible experimental demonstration of the entanglement and EPR steering between two magnon modes. To measure the quadratures of the magnon modes it is required to couple the magnons, via the beam-splitter-type interaction, to an additional weak microwave field with the dissipation rate significantly larger than that of the magnon modes. In this case, the magnon state can be read out via the output of the auxiliary microwave cavity mode, and the corresponding covariance matrix can be detected via balanced homodyne detection~\cite{Palomaki710,Vitali2007}.


\section{5 Conclusion}\label{SUMMARY}

In conclusion, we have shown that the presence of a cavity field can significantly enhance entanglement between two magnons in a two-sublattice ferrimagnet. The cases of antiferromagnet and ferrimagnet are studied in details. We have found that the enhancement of the entanglement is strong enough to generate two-way Einstein-Podolsky-Rosen (EPR) steering between the magnons. We have shown that in the presence of the cavity field a better entanglement and steering can be created in ferrimagnet than in antiferromagnet. We have found that in the absence of the cavity field no quantum steering can be generated when the magnons decay with the same rates. However, one-way steering then occurs when the magnons are damped with different rates. Moreover, the magnons are better steered in antiferromagnetic than in ferrimagnetic system. Detection of asymmetric EPR steering between the magnons is suggested as a way to determine the dissipation rates of the sublattices. The underlying physics is explained in terms of cooling of the magnon modes by the cavity photons, where the cavity plays the role of a cold bath which cools the magnons toward their vacuum state. We have also found an interesting interference effect in the system which destroys the enhancement of the entanglement and EPR steering by the creation of the perfect mutual coherence between one of the magnon and the cavity modes. It is found that the modes being mutually perfectly coherent are completely distinguishable. We would like to point out that our results contribute towards a better understanding of the generation and manipulation of quantum correlations in deep strong coupling regime. Since the magnons are macroscopic excitations of a magnet, the scheme discussed in this paper also provides a promising platform to study the macroscopic quantum phenomena such as the entanglement of massive objects and a possible way to test decoherence effects.

{\it This work is supported by the National Natural Science Foundation of China (Grants No. 11975026 and No. 61475007, and No. 61704071), the National Key R$\&$D Program of China (Grants No. 2018YFB1107200 and No. 2016YFA0301302), the Key R$\&$D Program of Guangzhou Province (Grant No. 2018B030329001), and Beijing Natural Science Foundation (Grant No. Z190005).}

\section*{Supplementary}
In this supplement, we give detailed derivation of the Hamiltonian (1) in the main text in Sec.~A, and in Sec.~B we derive analytical expressions for the steering parameters which show how the steering of the modes is related to the populations of the modes and their damping rates.

\subsection{A. The derivation of the Hamiltonian (1)}\label{quantization}
To quantize the Hamiltonian Eq.~(1) in the main text, we first introduce bosonic magnon operators 
by applying Holstein-Primakoff transfromation~\cite{Holstein1940}  
\begin{eqnarray}\label{HP_A}
S_{A,l}^{+}&=&\left( \sqrt{2S_A-a_l^\dagger a_l} \right )a_l,\nonumber \\
S_{A,l}^{-}&=&a_l^\dagger \sqrt{2S_A-a_l^\dagger a_l},\nonumber \\
S_{A,l}^{z}&=&S_A-a_l^\dagger a_l,
\end{eqnarray}
for spin operators on each site $l$ in sublattice $A$, and
\begin{eqnarray}\label{HP_B}
S_{B,l}^+&=&b_l^\dagger \sqrt{2S_B-b_l^\dagger b_l},\nonumber \\
S_{B,l}^-&=&\left( \sqrt{2S_B-b_l^\dagger b_l} \right) b_l,\nonumber \\
S_{B,l}^z&=&b_l^\dagger b_l-S_B,
\end{eqnarray}
for spin operators on the site $l$ in sublattice $B$. $S_{u,l}^{\pm}=S_{u,l}^{x}\pm iS_{u,l}^{y} (u=A,B)$ are the spin raising and lowering operators on site $l$. The introduced magnon operators $a_l, a_l^\dagger$ for the sublattice $A$ and $b_l, b_l^\dagger$ for the  sublattice $B$ satisfy the bosonic commutation relations. In the low-temperature limit, $\mathcal{H}_{\mathrm{FiM}}$ becomes
\begin{eqnarray}\label{H_FiM_site}
\mathcal{H}_{\mathrm{FiM}}&=&\omega_a\sum_l a_l^\dagger a_l+\omega_b\sum_l b_l^\dagger b_l  \nonumber \\
&&+2\sqrt{S_aS_b}J\sum_{l,\delta}(a_lb_{l+\delta}+a_l^\dagger b_{l+\delta}^\dagger),
\end{eqnarray}
where $\omega_i=\omega_{i,\mathrm{ex}}+\omega_{i,\mathrm{an}}+\omega_{i,0}~(i=a,b)$. The three terms composing the frequency $\omega_i$ originate from the exchange interaction between the sublattices $A$ and $B$, the anisotropy of the field and the presence of the external static field, respectively. The coefficients are $\omega_{a,\mathrm{ex}}=2ZS_BJ,~\omega_{b,\mathrm{ex}}=2ZS_AJ,~\omega_{a,\mathrm{an}}=g_A\mu_BH_{\mathrm{an},A},~\omega_{b,\mathrm{an}}=g_B\mu_BH_{\mathrm{an},B},~\omega_{a,0}=g_A\mu_BH,~\omega_{b,0}=-g_B\mu_BH$, with $Z$ denoting the coordination number.

We then introduce Fourier transformation,
\begin{eqnarray}
a_l&=&N^{-1/2}\sum_k e^{ik\cdot R_l}a_k,\nonumber \\
a_l^\dagger&=&N^{-1/2}\sum_k e^{-ik\cdot R_l}a_k^\dagger,\nonumber \\
b_l&=&N^{-1/2}\sum_k e^{ik\cdot R_l}b_k,\nonumber \\
b_l^\dagger&=&N^{-1/2}\sum_k e^{-ik\cdot R_l}b_k^\dagger,
\end{eqnarray}
to rewrite $\mathcal{H}_{\mathrm{FiM}}$ in momentum space as
\begin{equation}
\mathcal{H}_{\mathrm{FiM}}=\sum_k [ \omega_a a_k^\dagger a_k+\omega_b b_k^\dagger b_k+g_{ab}(a_k b_k+a_k^\dagger b_k^\dagger) ],
\end{equation}
where $N$ is the number of spins in each sublattice, $R_l$ is the vector pointing from origin to the $l$th site on the sublattice, and the coupling strength between the magnon modes is $g_{ab}=2\gamma_kZ\sqrt{S_A S_B}J$. $\gamma_k=Z^{-1}\sum_{\delta}e^{ik\delta}$ describes the structure factor. Here, $a_k,~a^\dagger_k$ ($b_k,~b^\dagger_k$) refer to the magnon annihilation and creation operators in momentum space with wave vector $k$ on the sublattice $A$ ($B$), meeting the commutation relations for bosons.

For the electromagnetic wave inside a cavity, the electric and magnetic field vectors in a finite volume $V$ take the following form
\begin{eqnarray}
\mathbf{E}_c&=&i\sqrt{\frac{\hbar \omega_k}{4\epsilon_0V}} \sum_k \left [ \mathbf{u}(\mathbf{k}) c_k e^{i\mathbf{k\cdot r}} - \mathbf{u}^*(\mathbf{k}) c_k^\dagger e^{-i\mathbf{k\cdot r}} \right ], \nonumber \\
\mathbf{H}_c&=&i\sqrt{\frac{\mu_0 \hbar \omega_k}{4V}} \sum_k \mathbf{k}\times \left [ \mathbf{u}(\mathbf{k}) c_k e^{i\mathbf{k\cdot r}} - \mathbf{u}^*(\mathbf{k}) c_k^\dagger e^{-i\mathbf{k\cdot r}} \right ], \nonumber \\
\end{eqnarray}
where $c_k^\dagger$ and $c_k$ are photon generation and annihilation operators in the momentum space that satisfy the Boson commutation relations, and $\mathbf{u}(k)$ is the polarization vector of the wave. By substituting the electric and magnetic fields into $\mathcal{H}_{c}$, we obtain the quantized Hamiltonian of the circularly polarized microwave
\begin{equation}
\mathcal{H}_{c} = \sum_k \omega_k c^\dagger_k c_k,
\end{equation}
with $\omega_k$ being the frequency of the photon mode. 

Taking the expressions of the quantized magnons and photons into account, the interaction Hamiltonian $\mathcal{H}_{\mathrm{int}}$ between the ferrimagnet and the microwave can be simplified as $\mathcal{H}_{\mathrm{int}}=\sum_k [g_{ac}(a_k c_k+a_k^\dagger c_k^\dagger)+g_{bc}(b_k c_k^\dagger+b_k^\dagger c_k)]$, where the coupling strength $g_{ic}=g_i\mu_B\sqrt{\mu_0\omega_cS_iN/2\hbar V}(i=A, B)$. Note that the interactions between the cavity with two types of magnons are different. The essential physics behind is that the spin orientation of sublattices $A$ and $B$ are opposite, i.e., the magnons excited on $A$ and $B$ have opposite angular momentum. One can see that the bosonic magnon operators $a$ and $b$ defined on two sublattices are different, as can be seen from Eqs.~(\ref{HP_A}) and (\ref{HP_B}). In the spin-up sublattice $A$, the effect of spin raising operator $S_{A}^+$ decreases the spin deviation from the equilibrium position. This corresponds to an annihilation process and is represented by the bosonic magnon annihilation operator $a$. In the spin-down sublattice $B$ the situation is different, the spin raising operator $S_{B}^+$ increases the spin deviation from the equilibrium position. This corresponds to a creation process and is represented by the bosonic magnon creation operator $b^\dagger$. Therefore, the two types of magnons are coupled with photons in different ways. Applying the Holstein-Primakoff transfromation, the coupling Hamiltonian $\sum_{i\in A,B}(c^\dagger S_i^{-}+cS_i^{+})$ takes the form $(c^\dagger b+c^\dagger a^\dagger + cb^\dagger+ca)$. Thus, photon mode $c$ couples to the magnon on sublattice $A$ through the parametric-type (two-mode squeezing) interaction, while it couples to the magnon on sublattice $B$ via the beam-splitter-type interaction.

Above all, we obtain the total Hamiltonian in the momentum space
\begin{align}
\mathcal{H}&=\sum_k  \left [\omega_a a_k^\dagger a_k+\omega_b b_k^\dagger b_k +g_{ab}(a_k^\dagger
b_k^\dagger + a_k b_k) \right ] \nonumber\\
& +  \sum_k \left [\omega_k c_k^\dagger c_k + g_{ac} \left ( a_k^ \dagger c_k^\dagger + a_k c_k \right )+ g_{bc}\left (b_k^\dagger c_k + b_k c_k^\dagger \right ) \right ].
\label{qm}
\end{align}
We consider the case of $k=0$ in this paper because photon can be strongly coupled with magnon only at $k=0$~\cite{YuanApl, Yuan2020PRB}. Then the resultant Hamiltonian reads
\begin{align}
\mathcal{H}&=  \omega_a a^\dagger a + \omega_b b^\dagger b +g_{ab}(a^\dagger
b^\dagger + a b) \nonumber\\
&\ \ \ \  + \omega_c c^\dagger c + g_{ac} \left ( a^ \dagger c^\dagger + a c \right)
+ g_{bc} \left (  b^\dagger c + b c^\dagger \right),
\end{align}
which is given in Eq.~(3) of the main text.

\subsection{B. The magnon-magnon steering in terms of populations}\label{steering criteria}

In this Appendix we demonstrate that the Gaussian steerability between the magnon modes is related to populations of the modes, and give analytical expressions which explicitly show that one-way steering occurs between the modes when the modes are unequally populated. Then, we derive analytical expressions for the steering parameters which explicitly show that in the absence of the cavity mode, one-way steering is possible when the modes are damped with different rates. 

In the present scheme, the reduced CM of modes $a$ and $b$ can be further expressed as
\begin{equation}
V'=\left(
\begin{array}{cccc}
n_a & 0 & c_1 & c_2 \\
0 & n_a & c_2 & -c_1 \\
c_1 & c_2 & n_b & 0 \\
c_2 & -c_1 & 0 & n_b
\end{array}
\right),
\end{equation}
where $V_{11}=V_{22}=n_a=\langle a^\dagger a\rangle+1/2$, $V_{33}=V_{44}=n_b=\langle b^\dagger b\rangle+1/2$, $V_{13}=-V_{24}=c_1$, $V_{14}=V_{23}=c_2$ with components $V_{ij}=\langle u_i (\infty)u_j (\infty)+ u_j(\infty) u_i (\infty)\rangle /2$ ($u=(X_a,Y_a,X_b,Y_b,X_c,Y_c)^{\mathrm{T}},~i,j=1,2,\ldots,6$) defined in the main text. Then the Gaussian steering parameters can be simplified to the following form
\begin{eqnarray} 
 \mathcal{G}^{a\rightarrow b}&=&\max{\{0,S(2V_a)-S(2V')\}} \nonumber \\
&=&\max{\{0,\frac{1}{2}\ln{\frac{\det{V_a}}{4\det{V'}}}\}} \nonumber \\
&=&\max{\{0,\frac{1}{2}\ln{\frac{n_a^2}{4\det{V'}}}\}}, \nonumber \\
 \mathcal{G}^{b\rightarrow a}&=&\max{\{0,S(2V_b)-S(2V')\}} \nonumber \\
 &=&\max{\{0,\frac{1}{2}\ln{\frac{\det{V_b}}{4\det{V'}}}\}} \nonumber \\
&=&\max{\{0,\frac{1}{2}\ln{\frac{n_b^2}{4\det{V'}}}\}},\label{B2}
 \end{eqnarray}
with $V_a=\left(\begin{array}{cc}n_a & 0  \\ 0 & n_a  \end{array} \right)$,~$V_b=\left(\begin{array}{cc}n_b & 0  \\ 0 & n_b  \end{array} \right)$. 

It is clearly seen from Eq.~(\ref{B2}) that when the populations of the modes $a$ and $b$ are equal, we can have either symmetric two-way steering or no steering. When $n_a>n_b$, we have three regimes of the parameters where different steering properties could be observed. Namely, when  $n_a^2>n_b^2>4\det V'$, two-way asymmetric steering with $\mathcal{G}^{a \rightarrow b}>\mathcal{G}^{b \rightarrow a}$ occurs; one-way steering with $\mathcal{G}^{a \rightarrow b}>\mathcal{G}^{b \rightarrow a}=0$ occurs when $n_a^2>4\det V'\geq n_b^2$, and no steering takes place when $4\det V'>n_a^2>n_b^2$.

In the absence of the cavity, the CM of the modes $a$ and $b$ can be significantly simplified. In this case, the elements of the matrix take the following form
\begin{eqnarray}
n_a&=&\frac{g_{ab}^2(\kappa_a+\kappa_b)\kappa_b}{De}+\frac{1}{2}, \nonumber \\
n_b&=&\frac{g_{ab}^2(\kappa_a+\kappa_b)\kappa_a}{De}+\frac{1}{2}, \nonumber \\
c_1&=&\frac{-g_{ab}\kappa_a\kappa_b(\omega_a+\omega_b)}{De}, \nonumber \\
c_2&=&\frac{-g_{ab}\kappa_a\kappa_b(\kappa_a+\kappa_b)}{De}, \nonumber \\
De&=&(\kappa_a\kappa_b-g_{ab}^2)(\kappa_a+\kappa_b)^2+\kappa_a\kappa_b(\omega_a+\omega_b)^2.
\end{eqnarray}

Then the steering in two directions become
\begin{eqnarray} \label{steering_OneChannel}
 \mathcal{G}^{a\rightarrow b}&=&\max{\{0,\frac{1}{2}\ln{\frac{n_a^2}{4\det{V'}}}\}}, \nonumber \\
 &=&\max{\{0,\ln{|1+\frac{2g_{ab}^2(\kappa_b-\kappa_a)\kappa_a}{De+2g_{ab}^2(\kappa_a^2+\kappa_b^2)}}|\}},\nonumber \\
 \mathcal{G}^{b\rightarrow a}&=&\max{\{0,\frac{1}{2}\ln{\frac{n_b^2}{4\det{V'}}}\}} \nonumber \\
 &=&\max{\{0,\ln{|1+\frac{2g_{ab}^2(\kappa_a-\kappa_b)\kappa_b}{De+2g_{ab}^2(\kappa_a^2+\kappa_b^2)}}|\}}.
 \end{eqnarray}
Clearly, in the case of $\kappa_a=\kappa_b$, no steering is observed, $\mathcal{G}^{a\rightarrow b}=\mathcal{G}^{b\rightarrow a}=0$. In other words, in the absence of the cavity mode, there is no steering between magnon modes which are damped with the same rates. When $\kappa_a\neq \kappa_b$, one-way steering occurs that the mode of smaller damping rate steers the mode of larger rate. For example, if $\kappa_a<\kappa_b$, then $\mathcal{G}^{a\rightarrow b}>0$ and $\mathcal{G}^{b\rightarrow a}=0$, i.e., one-way steering occurs in the direction $a\rightarrow b$.

\bibliography{ferrimagnetic_case}
\end{document}